\documentclass[aps,11pt,superscriptaddress,centertags,floatfix]{revtex4}%
\usepackage{graphicx}
\usepackage{amsmath}
\usepackage{amsfonts}
\usepackage{amssymb}%
\begin{document}
\title{Particle vibrational coupling in covariant density functional theory.
\footnote{This article is dedicated to Prof. Spartak Belyaev on the
occasion of his 85$^{\rm th}$ birthday}}
\author{\firstname{P.}~\surname{Ring}}
\email{ring@ph.tum.de}
\affiliation{Physik-Department der Technischen Universit\"at M\"unchen, D-85748 Garching, Germany}
\author{\firstname{E.}~\surname{Litvinova}}
\email{litvinova@gsi.de} \affiliation{GSI Helmholtzzentrum f\"{u}r
Schwerionenforschung, 64291 Darmstadt, Germany}
\affiliation{Frankfurt Institute for Advanced Studies,
Universit\"at Frankfurt, 60438 Frankfurt am Main, Germany}
\affiliation{Institute of Physics and Power Engineering, 249033
Obninsk, Russia}
\date{\today}

\begin{abstract}
A consistent combination of covariant density functional theory
(CDFT) and Landau-Migdal Theory of Finite Fermi Systems (TFFS) is
presented. Both methods are in principle exact, but Landau-Migdal
theory cannot describe ground state properties and density
functional theory does not take into account the energy dependence
of the self-energy and therefore fails to yield proper single-%
particle spectra as well as the coupling to complex configurations
in the width of giant resonances. Starting from an energy
functional, phonons and their vertices are calculated without any
further parameters. They form the basis of particle-vibrational
coupling leading to an energy dependence of the self-energy and an
induced energy-dependent interaction in the response equation. A
subtraction procedure avoids double counting. Applications in
doubly magic nuclei and in a chain of superfluid nuclei show
excellent agreement with experimental data.
\end{abstract}
\maketitle

\section{INTRODUCTION}

The understanding of the structure of nuclei far from stability with extreme
isospin is one the most exciting challenges of present nuclear physics. New
experimental facilities with radioactive nuclear beams make it possible to
investigate the nuclear chart to the very limits of nuclear binding. A wealth
of structure phenomena in exotic nuclei have been reported and the next
generation of radioactive-beam facilities will present new exciting
opportunities to study not only the ground states but also excitations and
spectra of these strongly interacting many-body systems. This situation has
stimulated considerable new efforts on the theoretical side to understand the
dynamics of the nuclear many-body problem by microscopic methods. Exact
solutions of the non-relativistic Schr\"{o}dinger equation based on the bare
nucleon-nucleon interaction are used to study very light nuclei with $A\leq12$
by an \textquotedblleft ab initio" approach, modern shell-model calculations
based on large scale diagonalization techniques and truncation schemes show
considerable success in situations where configuration mixing calculations are
possible, i.e. in light nuclei or in nuclei with single magic or doubly magic
configurations. For the large majority of nuclei, however, a quantitative
microscopic description is, so far, only possible by density functional theory
(DFT) and its extensions. Although DFT can, in principle, provide an exact
description of many-body problems\cite{KS.65}, if the exact density functional
is known, in nuclear physics one is far from a microscopic derivation of this
functional. In addition, nuclei are self-bound systems. As a consequence of
translational invariance the density in the laboratory frame is constant in
space. Density functional theory in nuclei is therefore based on the intrinsic
density, a concept that requires additional
approximations~\cite{Eng.07,Gir.08}. The most successful schemes of DFT in
nuclei use a phenomenological ansatz incorporating as many symmetries of the
system as possible and adjusting the parameters of the functional to ground
state properties of a few characteristic nuclei on the nuclear chart.
Considerable progress has been reported recently in constructing such
functionals. For a recent review see~\cite{BHR.03}.

One of the underlying symmetries of QCD is Lorentz invariance and therefore
covariant density functionals~\cite{Ring.96,VALR.05} are of particular
interest in nuclear physics. This symmetry not only allows to describe the
spin-orbit coupling, which has an essential influence on the underlying shell
structure, in a consistent way, but it also puts stringent restrictions on the
number of parameters in the corresponding functionals without reducing the
quality of the agreement with experimental data.

Most of the nuclei are superfluid systems and therefore the
inclusion of pairing correlations is essential for a correct
description of structure phenomena in open-shell
nuclei~\cite{Bel.61,Bel.68}. Hartree-Bogoliubov theory provides a
unified description of $ph$- and $pp$-correlations on a mean-field
level by using two type of densities, the normal density matrix
$\hat{\rho}=\langle a^{+}a\rangle$ and antisymmetric pairing
tensor $\hat{\kappa}=\langle aa\rangle$~\cite{RS.80}. According to
Valatin these two densities can be combined to the generalized
density matrix $\mathcal{\hat{R}}$ of double dimension
\cite{Val.61}. CDFT theory for superfluid systems is therefore
based on a generalized Relativistic Hartree-Bogoliubov (RHB)
energy density $E_{RHB}[\mathcal{\hat{R}}]$. The same is true for
the Landau-Migdal theory and for all the methods discussed in this
paper. For simplicity, however, we restrict all our considerations
in this article to the case without pairing correlations. Pairing
correlations can be included on all steps by using super-matrices.
Details are given in Ref. \cite{LRT.08}. Only in the applications
we present also calculations in isotopic chains of open shell
nuclei, that include pairing correlations.

A very successful example of a covariant density functional theory is the
Relativistic Hartree-Bogoliubov model~\cite{RHB}. It combines a density
dependence through a non-linear coupling between the meson fields~\cite{BB.77}
with pairing correlations based on an effective interaction of finite range. A
large variety of nuclear phenomena have been described over the years within
this model: the equation of state in symmetric and asymmetric nuclear matter,
ground state properties of finite spherical and deformed nuclei all over the
periodic table~\cite{GRT.90} from light nuclei~\cite{LVR.04a} to super-heavy
elements~\cite{LSRG.96}, from the neutron drip line, where halo phenomena are
observed~\cite{MR.96} to the proton drip line~\cite{LVR.04} with nuclei
unstable against the emission of protons~\cite{LVR.99}.

In principle density functional theory can be used for the
description of all properties depending on the single-particle
density. It is therefore not only limited to the description of
the ground state properties. The same density functionals have
also been applied for a very successful description of excited
states, such as rotational bands in normal and super-deformed
nuclei~\cite{AKRE.00,ARK.00} and collective
vibrations~\cite{MWG.02}. Rotations are treated in the cranking
approximation providing a quasi-static description of the nuclear
dynamics in a rotating frame and for the description of vibrations
a time-dependent mean field approximation is used by assuming
independent particle motion in time-dependent average
fields~\cite{VBR.95}. In the small amplitude limit one obtains the
relativistic Random Phase Approximation (RRPA)~\cite{RMG.01} and
in superfluid nuclei the relativistic Quasiparticle Random Phase
Approximation (RQRPA)~\cite{PRN.03}. This method provides a
natural framework to investigate collective and non-collective
excitations of $ph$- (or $2qp$) character. It is successful in
particular for the understanding of the position of giant
resonances and spin- or/and isospin-excitations as the Gamov
Teller Resonance (GTR) or the Isobaric Analog Resonance (IAR).
Recently it has been also used for a theoretical interpretation of
low lying E1-strengths observed in neutron rich isotopes (pygmy
modes)~\cite{PRN.03} and for low-lying collective quadrupole
excitations~\cite{Ans.05}.

Density functional theory in nuclei is based on intrinsic
densities and on the mean field approach. Therefore it cannot
provide an exact treatment of the full nuclear dynamics. It breaks
down in \ transitional nuclei, where the intrinsic density is not
well defined and where one has to include correlations going
beyond the mean field approximation by treating quantum
fluctuations through a superposition of several mean field
solutions, as for instance in the Generator Coordinate Method
(GCM)~\cite{RS.80}. But it also provides a poor approximation for
the single-particle spectra particularly in ideal shell model
nuclei such as $^{208}$Pb with closed protons and neutron shells.
One finds in self-consistent mean field calculations usually a
considerably enhanced Hartree-Fock gap in the single-particle
spectrum and a reduced level density at the Fermi surface as
compared with the experiment. It is well known that this fact is
connected with the relatively small effective mass in such models.
Mahaux and collaborators~\cite{JLM.76} have shown that the
effective mass in nuclear matter is roughly $m^{\ast}/m$
$\approx0.8$. In finite nuclei it should be modified by the
coupling of the single-particle motion to low-lying collective
surface vibrations. This leads, in the vicinity of the Fermi
surface, to an enhancement of $m^{\ast}/m$ $\approx1$.
Non-self-consistent models with the bare mass ( $m^{\ast}/m$
$\approx1$) show indeed a single-particle spectrum with a level
density close to the experiment. With a few exceptions, where the
quadrupole motion has been studied within the relativistic
Generator Coordinate Method (GCM)~\cite{NVR.06a,NVR.06b},
applications of covariant density functional theory to the
description of excited states are limited to relativistic RPA,
i.e. to configurations of $1p1h$-nature. None of these methods,
however, can be applied to the investigation of the coupling to
more complicated configurations, as it occurs for instance in the
damping phenomena causing the width of giant resonances.

Already before density functional theory has been introduced in the sixties
for the description of quantum mechanical many-body problems by Kohn and
Sham~\cite{KS.65} Landau has developed in the fifties his Fermi Liquid Theory
(FLT)~\cite{Lan.59} for infinite systems. It has been extended to the Theory
of Finite Fermi Systems (TFFS)~\cite{Mig.67} by Migdal. This theory provides
another very successful method for the description of low-lying nuclear
excitations~\cite{RSp.74}. It has several general properties in common with
density functional theory. First, both theories are know to be exact, at least
in principle, but in practice, in nuclear physics, the parameters entering
these theories have to be determined in a phenomenological way by adjustment
to experimental data. Second, both theories are based on a single-particle
concept. DFT uses the mean field concept with Slater determinants in an
effective single-particle potential as a vehicle to introduce shell-effects in
the exact density functional introduced by Hohenberg and Kohn~\cite{HK.64}.
Fermi liquid theory is based on the concept of quasi-particles obeying a Dyson
equation, which are defined as the basic excitations of the neighboring system
with odd particle number. Third, in practical applications both theories
describe in the simplest versions the nuclear excitations in the RPA
approximation, i.e. by a linear combination of $ph$-configurations in an
average nuclear potential.

However, there are also essential differences between these two
concepts. First, in contrast to DFT, TFFS does not attempt to
calculate the ground state properties of the many-body system, but
it describes the nuclear excitations in terms of Landau
quasi-particles and their interaction. Therefore the experimental
data used to fix the phenomenological parameters of the theories
are bulk properties of the ground state in the case of density
functional theory, and properties of single-particle excitations
and of the collective excitations in the case of finite Fermi
systems theory. Second, in DFT the mean field is determined in a
self-consistent way and therefore the RPA spectrum contains
Goldstone modes at zero energy. This is usually not the case in
TFFS calculations, which are based on a non-relativistic
shell-model potential, whose parameters are adjusted to the
experimental single-particle spectra. Therefore, apart from a few
approximate attempts to treat the Goldstone modes by adjusting
additional parameters in the effective quasiparticle interaction,
there is no self-consistency in the RPA calculations of TFFS and
the Goldstone modes do not separate from the other modes. They are
distributed among the low-lying excitations. Third, modern
versions of TFFS go much beyond the mean field approximation. The
coupling between the particles and the phonons is investigated
with Green's function techniques. Based on the phonons calculated
in the framework of the RPA one has included particle-phonon
coupling vertices and an energy dependence of the self-energy in
the Dyson equation~\cite{RW.73,HS.76}. This leads also to an
induced interaction in the Bethe-Salpeter equation caused by the
exchange of phonons which also depends on the energy. The coupling
of particles and phonons has also been derived from Nuclear Field
Theory (NFT) and its extensions~\cite{BM.75,BBB.77,BBB.83}. Many
aspects of the coupling between the quasi-particles and the
collective vibrations have been investigated with these
techniques~\cite{BBBD.79,BB.81,CBG.92,CB.01,SBC.04,Tse.89,KTT.97,KST.04,Tse.05,LT.05}
as well as with other kinds of approaches beyond
RPA~\cite{Sol.92,DNSW.90} over the years.

We give here an overview over recent
attempts~\cite{LR.06,LRT.07,LRT.08,LRT.09} to find a combination
of the basic ideas of covariant density functional theory and
Landau-Migdal theory and show as examples corresponding
applications. The concept is similar to earlier work in
Refs.~\cite{KS.79,KS.80,FTT.00}, where specific non-relativistic
energy functionals have been used to construct a Self-Consistent
Theory of Finite Fermi Systems. The starting point is a covariant
density functional $E[\rho]$ widely used in the literature. It is
adjusted to ground state properties of characteristic nuclei and,
without any additional parameter, it provides the necessary input
of finite Fermi systems theory, such as the mean field and the
single-particle spectrum as well as an effective interaction
between the $ph$-configurations in terms of the second derivative
of the same energy $E[\rho]$ with respect to the density. Thus the
phenomenological input of Landau-Migdal theory is replaced by the
results of density functional theory. The same interaction is used
to calculate the vertices for particle-vibration
coupling~\cite{LR.06}. In a second step techniques of
Landau-Migdal theory and its modern extensions are used to
describe the coupling of one- and two-quasiparticle
configurations. The main assumption of the quasiparticle-phonon
coupling model is that two types of elementary excitations --
two-quasiparticle and vibrational modes -- are coupled in such a
way that configurations of $1p1h\otimes phonon$ type with
low-lying phonons strongly compete with simple $1p1h$
configurations close in energy or, in other words, that
quasiparticles can emit and absorb phonons with rather high
probabilities. In this way a fully consistent description of the
many-body dynamics is obtained. As a result an induced additional
interaction between single-particle and vibrational excitations
provides a strong fragmentation of the pure RQRPA states causing
the spreading width of giant resonances and the redistribution of
the pygmy strength.

Two essential approximations are used in this context: First, the
\textit{Time-Blocking Approximation} (TBA)~\cite{Tse.89}, that has
been extended to systems with pairing correlations (QTBA) in Ref.
~\cite{LRT.08}, blocks in a special time-projection technique the
$1p1h$-propagation through states which have a more complex
structure than $1p1h\otimes phonon$. The nuclear response can then
be explicitly calculated on the $1p1h+1p1h\otimes phonon$ level by
summation of an infinite series of Feynman's diagrams. Second, a
special \textit{subtraction technique} guarantees, that there is
no double counting between the additional correlations introduced
by particle-vibration coupling and the ground state correlations
already taken into account in the phenomenological density
functional. These two tools are essential for the success of this
method. TBA introduces a consistent truncation scheme into the
Bethe-Salpeter equation and without it it would be hard to solve
the equations explicitly. The subtraction method is the essential
tool to connect density functional theory so far used only on the
level of mean field theory, i.e. on the RPA level, with the
extended Landau-Migdal theory, where complex configurations are
included through particle-vibration coupling.

The structure of the paper is the following: In Section II we
discuss shortly the general formalism of covariant density
functional theory, we introduce in Section III the concept of the
energy-dependent self-energy $\Sigma (\varepsilon)$ and the
vertices of particle-vibration coupling in the relativistic
framework, and we discuss in Section IV the response formalism,
the time blocking approximation and the subtraction mechanism for
the response function. In Section V we present recent numerical
applications for the calculation of level densities at the Fermi
surface and the spreading width of the several nuclei. Section VI
contains a brief summary and an outlook for future applications.

\section{THE RELATIVISTIC ENERGY DENSITY FUNCTIONAL}

Covariant density functional theory uses the Walecka
model~\cite{SW.86} as a Lorentz invariant framework for the
formulation of the density functional. In this model the nucleus is
described as a system of Dirac nucleons coupled to the exchange
mesons and the electromagnetic field through an effective Lagrangian.
The isoscalar scalar $\sigma$-meson, the isoscalar vector
$\omega$-meson, and the isovector vector $\rho$-meson build the
minimal set of meson fields that together with the electromagnetic
field is necessary for a quantitative description of bulk and
single-particle nuclear
properties~\cite{SW.86,SW.97,Rei.89,Ser.92,Ring.96}. The model is
defined by the Lagrangian density
\begin{equation}
\mathcal{L}=\mathcal{L}_{N}+\mathcal{L}_{m}+\mathcal{L}_{int}%
.\label{Lagrangian}%
\end{equation}
$\mathcal{L}_{N}$ denotes the Lagrangian of the free nucleon
\begin{equation}
\mathcal{L}_{N}=\bar{\psi}\left(  i\gamma^{\mu}\partial_{\mu}-m\right)  \psi,
\end{equation}
where $m$ is the bare nucleon mass and $\psi$ denotes the Dirac spinor.
$\mathcal{L}_{m}$ is the Lagrangian of the free meson fields and the
electromagnetic field
\begin{align}
\mathcal{L}_{m} &  =\frac{1}{2}\partial_{\mu}\sigma\partial^{\mu}\sigma
-\frac{1}{2}m_{\sigma}^{2}\sigma^{2}-\frac{1}{4}\Omega_{\mu\nu}\Omega^{\mu\nu
}+\frac{1}{2}m_{\omega}^{2}\omega_{\mu}\omega^{\mu}\nonumber\\
&  -\frac{1}{4}\vec{R}_{\mu\nu}\vec{R}^{\mu\nu}+\frac{1}{2}m_{\rho}^{2}%
\vec{\rho}_{\mu}\vec{\rho}^{\mu}-\frac{1}{4}F_{\mu\nu}F^{\mu\nu},
\end{align}
with the corresponding masses $m_{\sigma}$, $m_{\omega}$,
$m_{\rho}$, and $\Omega_{\mu\nu}$, $\vec{R}_{\mu\nu}$,
$F_{\mu\nu}$ are field tensors (arrows denote isovectors and
boldface symbols are for vectors in ordinary space). The minimal
set of interaction terms is contained in $\mathcal{L}_{int}$
\begin{equation}
\mathcal{L}_{int}=-\bar{\psi}\Gamma_{\sigma}\sigma\psi-\bar{\psi}%
\Gamma_{\omega}^{\mu}\omega_{\mu}\psi-\bar{\psi}\vec{\Gamma}_{\rho}^{\mu}%
\vec{\rho}_{\mu}\psi-\bar{\psi}\Gamma_{e}^{\mu}A_{\mu}\psi.
\end{equation}
with the vertices%
\begin{equation}
\Gamma_{\sigma}=g_{\sigma},\;\;\;\Gamma_{\omega}^{\mu}=g_{\omega}\gamma^{\mu
},\;\;\;\vec{\Gamma}_{\rho}^{\mu}=g_{\rho}\vec{\tau}\gamma^{\mu}%
,\;\;\;\;\Gamma_{e}^{\mu}=q\gamma^{\mu},
\end{equation}
with the coupling constants $g_{\sigma}$, $g_{\omega}$, $g_{\rho}$
and $q$ ($e$ or $0$ for protons or neutrons). Already in the
earliest applications of the RMF framework it was realized,
however, that this simple model with interaction terms only linear
in the meson fields, does not provide a quantitative description
of complex nuclear systems. An effective density dependence was
introduced~\cite{BB.77} by replacing the quadratic $\sigma
$-potential $\frac{1}{2}m_{\sigma}^{2}\sigma^{2}$ with a
non-linear meson coupling potential \thinspace$U(\sigma)$, which
contains additional parameters. This particular form of the
non-linear potential has become standard in applications of RMF
models, although additional non-linear interaction terms, both in
the isoscalar and isovector channels, have also been
considered~\cite{Bod.91,ST.94,SFM.00,HP.01}.

>From the model Lagrangian density the classical variation principle leads to
the equations of motion:%
\begin{equation}
\left[  \gamma^{\mu}(i\mathbf{\partial}_{\mu}+V_{\mu})+m+S\right]  \psi=0.
\label{Dirac0}%
\end{equation}
If one neglects retardation effects for the meson fields, which is well
justified because of the large meson masses, a self-consistent solution is
obtained when the time-dependent mean-field potentials
\begin{equation}
S=g_{\sigma}\sigma~,\text{ \ \ }V_{\mu}=g_{\omega}\omega_{\mu}+g_{\rho}%
\vec{\tau}\vec{\rho}_{\mu}+qA_{\mu}~,\nonumber
\end{equation}
are calculated at each step in time by the solution of the static Klein-Gordon
equations
\begin{equation}
-\Delta\phi_{m}+U^{\prime}(\phi_{m})=\pm\left\langle \bar{\psi}\Gamma_{m}%
\psi\right\rangle , \label{KG0}%
\end{equation}
where the ($+$) sign is for vector fields and the ($-$) sign for the scalar
field. The index $m$ denotes mesons and the photon, i.e. $\phi_{m}%
\equiv\{\sigma,\omega^{\mu},\vec{\rho}^{\mu},A^{\mu}\}$, and $U^{\prime}%
(\phi_{m})$ is derivative of the corresponding potential with respect to the
meson field.

In applications to nuclear matter and finite nuclei, the relativistic models
are used in the \textit{no-sea} approximation, i.e. the Dirac sea of states
with negative energies does not contribute to the densities and currents and
one uses
\begin{equation}
\left\langle \bar{\psi}\Gamma_{m}\psi\right\rangle =\sum\limits_{i=1}^{A}%
\bar{\psi}_{i}^{{}}(\mathbf{r},t)\Gamma_{m}\psi_{i}^{{}}(\mathbf{r},t)\;,
\label{Densities}%
\end{equation}
where the sum runs only over the occupied states in the Fermi sea, i.e. vacuum
polarization effects are neglected. In fact, many effects that go beyond the
classical mean-field level are apparently neglected in the this models: Fock
terms, vacuum polarization effects, and the short range Brueckner-type
correlations. The experimental data to which the meson-nucleon couplings are
adjusted, however, contain all these effects and much more. It follows that
effects beyond the mean-field level are implicitly included in the RMF
approach by adjusting the model parameters to reproduce a selected empirical
data set. Vacuum effects, chiral symmetry, nucleon substructure, exchange
terms, long- and short-range correlation effects are, therefore, effectively
included in this approach although neither of them can be accessed separately.

The set of coupled equations (\ref{Dirac0}) and (\ref{KG0}) define the
relativistic mean field (RMF) model. In the stationary case they reduce to a
nonlinear eigenvalue problem and in the time-dependent case they describe the
nonlinear propagation of the Dirac spinors in time~\cite{VBR.95}.

RMF models can be also formulated without explicitly including mesonic degrees
of freedom. Meson-exchange interactions can be replaced by local four-point
interactions between nucleons. It has been shown that the relativistic
point-coupling models~\cite{MM.89,BMM.02,NVLR.08} are completely equivalent to
the standard meson-exchange approach. In order to describe properties of
finite nuclei on a quantitative level, the point-coupling models include also
some higher order interaction terms. For instance, six-nucleon vertices
$(\bar{\psi}\psi)^{3}$, and eight-nucleon vertices $(\bar{\psi}\psi)^{4}$ and
$[(\bar{\psi}\gamma_{\mu}\psi)(\bar{\psi}\gamma^{\mu}\psi)]^{2}$.

These relatively simple models turn out to provide a very successful
phenomenological description of the nuclear many-body system all over the
periodic table. Relatively few parameters are adjusted to ground state
properties of a few finite nuclei. At a first glance it is not easy to see how
can such a simple approach can be so successful. This can be only understood
if one considers that this model represents an approximate implementation of
Kohn-Sham density functional theory (DFT)\cite{KS.65,Koh.99,DG.90}, which is
successfully employed in the treatment of the quantum many-body problem in
atomic, molecular and condensed matter physics.

It is evident that equations of motion (\ref{Dirac0}) and (\ref{KG0}) can also
be directly derived form a density functional. Using the definition of the
relativistic single-nucleon density matrix
\begin{equation}
\hat{\rho}(\mathbf{r,r}^{\prime},t)=\sum\limits_{i=1}^{A}|\psi_{i}%
(\mathbf{r,}t)\rangle\langle\psi_{i}(\mathbf{r}^{\prime}%
,t)|\;,\label{density-HF}%
\end{equation}
the total energy can be written as a functional of the density matrix
$\hat{\rho}$ and of the meson fields
\begin{equation}
E_{RMF}[\hat{\rho},\phi_{m}]=\text{Tr}\left[ ({\mbox{\boldmath
$\alpha$}}\mathbf{p}+\beta
m)\hat{\rho}\right]  \pm\int\left[  \frac{1}{2}(\mathbf{\nabla}\phi_{m}%
)^{2}+U(\phi_{m})\right]  d^{3}r+\text{Tr}\left[  (\Gamma_{m}\phi_{m}%
)\hat{\rho}\right]  .\label{E-RMF}%
\end{equation}
The trace operation involves a sum over the Dirac indices and an integral in
coordinate space. The index $m$ is used as generic notation for all mesons and
the photon.

\section{THE ENERGY DEPENDENCE OF THE SELF-ENERGY}

In a relativistic many-body theory the motion of single nucleons in the
nuclear medium is described by the Dyson equation
\begin{equation}
\bigl(\gamma^{\mu}P_{\mu}-m^{\ast}\bigr)|\psi\rangle=0,
\end{equation}
where the self-energy is given by
\begin{equation}
m^{\ast}=m+\Sigma_{s}\label{emass}%
\end{equation}
with the scalar part $\Sigma_{s}$ of the self-energy and where the
generalized four-vector momentum operator has the form
\begin{equation}
P_{\mu}=p_{\mu}-\Sigma_{\mu}=\Bigl(i\frac{\partial}{\partial t}-\Sigma
_{0},i\mathbf{\nabla}+\mathbf{\Sigma}\Bigr)
\end{equation}
with the vector part $\Sigma^{\mu}$ of the self-energy
\begin{equation}
\Sigma^{\mu}=(\Sigma^{0},\mathbf{\Sigma}).
\end{equation}
The index '$s$' in the Eq. (\ref{emass}) denotes that the effective mass is
described by the scalar $\sigma$-meson field. In order to characterize ground
state properties the stationary Dirac equation has to be solved:
\begin{equation}
\bigl({\mbox{\boldmath $\alpha$}}(\mathbf{p}-\mathbf{\Sigma})+\beta m^{\ast
}+\Sigma_{0}\bigr)|\psi\rangle=\varepsilon|\psi\rangle.\label{Deq1}%
\end{equation}
In the general case the full self-energy is non-local in space and
also in time. This non-locality means that its Fourier transform
has both momentum and energy dependence. We therefore decompose
the total self-energy in a stationary local part and an energy-%
dependent non-local term:
\begin{equation}
\Sigma(\mathbf{r},\mathbf{r^{\prime}};\omega)={\tilde{\Sigma}}(\mathbf{r}%
)\delta(\mathbf{r}-\mathbf{r^{\prime}})+\Sigma^{e}(\mathbf{r}%
,\mathbf{r^{\prime}};\omega),\label{sig}%
\end{equation}
where all the components of the self-energy are involved:
\begin{equation}
{\Sigma}=({\Sigma}_{s},{\Sigma}_{\mu}),\text{ \ \ \ \ \ \ }\tilde{\Sigma
}=(\tilde{\Sigma}_{s},\tilde{\Sigma}_{\mu}),\text{ \ \ \ \ \ }{\Sigma}%
^{e}=({\Sigma}_{s}^{e},{\Sigma}_{\mu}^{e})\nonumber
\end{equation}
and the index "e" indicates the energy dependence.

The energy-independent parts of the self-energy correspond to the
average fields of the Walecka model:
\begin{equation}
{\tilde{\Sigma}}_{s}(\mathbf{r})=S(\mathbf{r)},\text{ \ \ \ \ }{\tilde{\Sigma
}}_{\mu}(\mathbf{r})=V_{\mu}(\mathbf{r})
\end{equation}
These fields satisfy the inhomogeneous Klein-Gordon equations, where the
sources are determined by the respective density and current distributions in
a system of A nucleons.

We assume time-reversal symmetry that means the absence of currents in the
nucleus and, thus, we find vanishing space-like components of $\ {\Sigma}$.
The equation of the one-nucleon motion has the form:
\begin{equation}
\bigl(h_{D}+\beta\Sigma_{s}^{e}(\varepsilon)+\Sigma_{0}^{e}(\varepsilon
)\bigr)|\psi\rangle=\varepsilon|\psi\rangle
\end{equation}
where $h_{D}$ is the Dirac Hamiltonian
\begin{equation}
h_{D}={\mbox{\boldmath $\alpha$}}\mathbf{p}+\beta(m+{\tilde{\Sigma}}%
_{s})+{\tilde{\Sigma}}_{0} \label{Dirac-H}%
\end{equation}
\textbf{ }or, in the language of Green's functions
\begin{equation}
\bigl(\varepsilon-h_{D}-\beta\Sigma_{s}^{e}(\varepsilon)-\Sigma_{0}%
^{e}(\varepsilon)\bigr)G(\varepsilon)=1. \label{fg}%
\end{equation}
It turns out to be useful to work in the shell-model Dirac basis $\{|\psi
_{k}\rangle\}$ which diagonalizes the energy-independent part of the Dirac
equation:
\begin{equation}
h_{D}|\psi_{k}\rangle=\varepsilon_{k}|\psi_{k}\rangle. \label{dirac-basis}%
\end{equation}
In this basis one can rewrite Eq. (\ref{fg}) as follows:
\begin{equation}
\sum\limits_{l}\bigl\{(\varepsilon-\varepsilon_{k})\delta_{kl}-\Sigma_{kl}%
^{e}(\varepsilon)\bigr\}G_{lk^{\prime}}(\varepsilon)=\delta_{kk^{\prime}},
\label{mo}%
\end{equation}
where the letter indices $k,k^{\prime},l$ denote full sets of the
spherical quantum numbers.

\begin{figure}[ptb]
\begin{center}
\includegraphics*[scale=0.8]{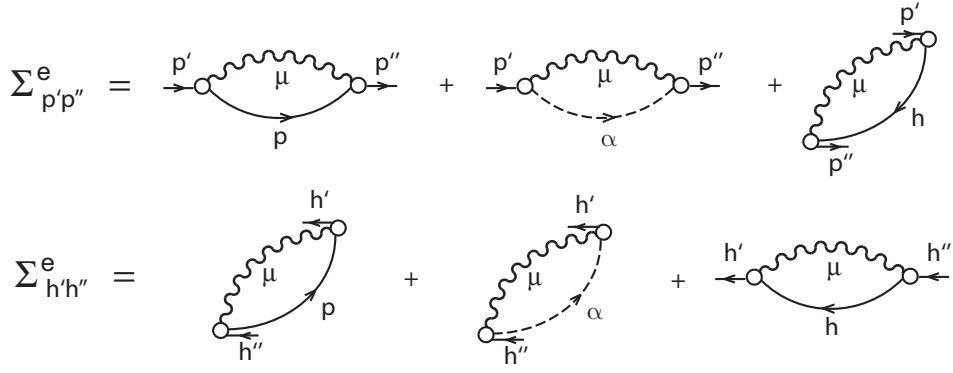}
\end{center}
\vspace{-1cm}%
\caption{The particle
$\Sigma^{e}_{p^{\prime}p^{\prime\prime}}$ and the hole
$\Sigma^{e}_{h^{\prime}h^{\prime\prime}}$ components of the
relativistic self-energy in the graphical representation. Solid and
dashed lines with arrows denote one-body propagators for particle
($p$), hole ($h$), and antiparticle ($\alpha$) states. Wavy lines
denote phonon ($\mu$) propagators, empty circles
are the particle-phonon coupling vertices $\gamma^{\mu}$ in Eq. (\ref{phonon}%
). The time direction is from the left to the right.}%
\label{fig1}%
\end{figure}

Obviously, on this stage one needs some model assumptions. The particle-phonon
coupling model~\cite{BM.75} provides a rather simple approximation to describe
the energy dependence of $\Sigma^{e}(\varepsilon)$. Within this model
$\Sigma^{e}(\varepsilon)$ is a convolution of the particle-phonon coupling
amplitude $\Gamma$ and the exact single-particle Green's function~\cite{KT.86}%
:
\begin{equation}
\Sigma_{kl}^{e}(\varepsilon)=\sum\limits_{k^{\prime}l^{\prime}}\int
\limits_{-\infty}^{+\infty}\frac{d\omega}{2\pi i}\Gamma_{kl^{\prime}%
lk^{\prime}}(\omega)G_{k^{\prime}l^{\prime}}(\varepsilon+\omega),\label{mo1}%
\end{equation}
where the amplitude
\begin{equation}
\Gamma_{kl^{\prime}lk^{\prime}}(\omega)=-\sum\limits_{\mu}\Bigl(\frac
{\gamma_{k^{\prime}k}^{\mu\ast}\gamma_{l^{\prime}l}^{\mu}}{\omega-\Omega^{\mu
}+i\eta}-\frac{\gamma_{kk^{\prime}}^{\mu}\gamma_{ll^{\prime}}^{\mu\ast}%
}{\omega+\Omega^{\mu}-i\eta}\Bigr)
\end{equation}
is represented in terms of phonon vertexes $\gamma^{\mu}$ and their
frequencies $\Omega^{\mu}$. They are determined by the following relation:
\begin{equation}
\gamma_{kl}^{\mu}=\sum\limits_{k^{\prime}l^{\prime}}\tilde{V}_{kl^{\prime
}lk^{\prime}}\delta\rho_{k^{\prime}l^{\prime}}.\label{phonon}%
\end{equation}
$\delta\rho$ is the transition density and in the linearized version of the
model $\tilde{V}_{kl^{\prime}lk^{\prime}}$ denotes the relativistic matrix
element of the static residual interaction, i.e. the second derivative of the
energy functional with respect to the density matrix
\begin{equation}
\tilde{V}=\frac{\delta^{2}E_{RMF}[\hat{\rho}]}{\delta\hat{\rho}\delta\hat
{\rho}}\label{Vstatic}%
\end{equation}
In the linear approximation $\delta\rho$ is not influenced by the
particle-phonon coupling and can be computed within relativistic
RPA or QRPA. The linearized version implies also that the
energy-dependent part of the self-energy (\ref{mo1}) contains the
mean field Green's function
${\tilde{G}}(\varepsilon)=(\varepsilon-h_{D})^{-1}$ instead of the
exact Green's function $G$. The graphical representation of the
self-energy is given in Fig.~\ref{fig1}. In contrast to the
non-relativistic case, where one has occupied states below the
Fermi surface (hole states $h$) and empty states above the Fermi
surface (particle states $p$) we now have according to the no-sea
approximation in addition empty states with negative energies in
the Dirac sea (anti-particle states $\alpha$). Particle and hole
components are drawn assuming all the possible types of
intermediate states.

Eq. (\ref{mo}) contains off-diagonal elements of the matrix $\Sigma_{kl}^{e}$
with relatively large energy denominators. It has been shown by explicit
calculations within the non-relativistic approach~\cite{RW.73} that it is
justified to use the diagonal approximation:
\begin{equation}
\Sigma_{kl}^{e}(\varepsilon)=\delta_{kl}\Sigma_{k}^e(\varepsilon).
\label{diagonal}%
\end{equation}
Thus, within the diagonal approximation of the self-energy
(\ref{diagonal}) the exact Green's function $G$ is also diagonal
in the Dirac basis and the Dyson equation forms for each $k$ a
non-linear eigenvalue equation
\begin{equation}
(\varepsilon-\varepsilon_{k}-\Sigma_{k}^{e}(\varepsilon))G_{k}(\varepsilon)=1.
\label{Dyson}%
\end{equation}
The poles of the Green's function $G_{k}(\varepsilon)$ correspond to the zeros
of the function
\begin{equation}
f(\varepsilon)=\varepsilon-\varepsilon_{k}-\Sigma_{k}^{e}(\varepsilon).
\label{zeros}%
\end{equation}
In Refs. \cite{RW.73,LR.06} it is shown how this problem can be
solved by a matrix diagonalization. For each quantum number $k$
there exist several solutions $\varepsilon_{k}^{(\lambda)}$
characterized by the index $\lambda$. Because of the coupling to
the collective vibrations the single-particle state $k$ is
fragmented as it will be shown in the application in section
\ref{appl-1}.

\section{THE RESPONSE FUNCTION}

\begin{figure}[ptb]
\begin{center}
\includegraphics*[scale=0.75]{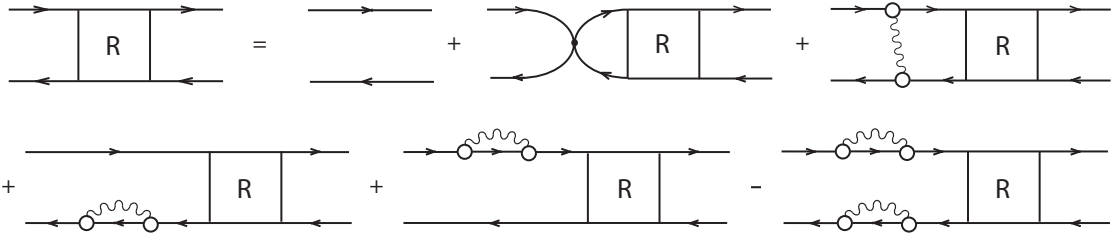}
\end{center}
\vspace{-0.1cm}%
\caption{Bethe-Salpeter equation for the $ph$-response function $R$
in graphical representation. Details are given in Fig.~\ref{fig1} and
the small black circle means the static part of the residual
$ph$-interaction
(\ref{Vstatic}). }%
\label{fig2}%
\end{figure}

Nuclear dynamics of an even-even nucleus in a weak external field is described
by the linear response function $R(14,23)$, where $1=\{k_{1},t_{1}\}$ combines
the quantum numbers $k$ and the time. This response function is the solution
of the Bethe-Salpeter equation (BSE) in the $ph$ channel:
\begin{equation}
R(14,23)=R^{0}(14,23)-i\sum\limits_{5678}R^{0}%
(16,25)V(58,67)R(74,83),\label{bse}%
\end{equation}
with the free response $R^{0}(14,23)=G(1,3)G(4,2)$, where function $G$ is the
exact single-particle Green's function. The summation over the number indices
$1$, $2,\dots$ implies an integration over the respective time variables. and
$V$ is the effective interaction irreducible in the $ph$-channel. This
interaction is determined as a variational derivative of the full self-energy
$\Sigma$ with respect to the exact single-particle Green's function:
\begin{equation}
V(14,23)=i\frac{\delta\Sigma(4,3)}{\delta G(2,1)}.
\end{equation}
Since the self-energy in Eq.~(\ref{sig}) has two parts
$\Sigma=\tilde{\Sigma }+\Sigma^{e}$, the effective interaction $V$
in Eq.~(\ref{bse}) is a sum of the static RMF interaction
$\tilde{V}$ (\ref{Vstatic}) and time-dependent terms
\begin{equation}
V^{e}(14,23)=i\frac{\delta{\Sigma^{e}(4,3)}}{\delta G(2,1)}\label{dcons}%
\end{equation}
After a Fourier transformation in time, this time dependence leads
to an energy-dependent interaction $V^{e}$. In the Dirac basis \
(\ref{dirac-basis}) it has the form:
\begin{equation}
V_{kl^{\prime},lk^{\prime}}^{e}(\omega,\varepsilon,\varepsilon^{\prime}%
)=\sum\limits_{\mu,\sigma}\frac{\sigma\gamma_{k^{\prime}k}^{\mu(\sigma)\ast
}\gamma_{l^{\prime}l}^{\mu(\sigma)}}{\varepsilon-\varepsilon^{\prime}%
+\sigma(\Omega^{\mu}-i\eta)}.
\end{equation}
where $\sigma=+1$ for empty states and $-1$ for occupied states
(for details see Ref. \cite{LRT.07}). The Bethe-Salpeter equation
(\ref{bse}) contains the exact Greens' function $G$. \ In order to
simplify this equation for the further analysis the $G$ is
expressed it in terms of the mean field Green's function
$\tilde{G}$. From Eq. (\ref{mo}) we derive the Nambu form for it:
\begin{equation}
{\tilde{G}}^{-1}(1,2)=G^{-1}(1,2)+\Sigma^{e}(1,2),
\end{equation}
which reads in Fourier space as%
\begin{equation}
{\tilde{G}}_{k_{1}k_{2}}(\varepsilon)=\frac{\delta_{k_{1}k_{2}}}%
{\varepsilon-\varepsilon_{k_{1}}+i\sigma_{k_{1}}\eta}.\label{G-tilde}%
\end{equation}
Introducing ${\tilde{R}}^{0}(14,23)={\tilde{G}}(1,3){\tilde{G}}(4,2)$ one can
rewrite Eq.~(\ref{bse}) as follows:
\begin{equation}
R={\tilde{R}}^{0}-i\tilde{R}^{0}WR,\label{bse1}%
\end{equation}
where $W$ is a new interaction of the form
\begin{equation}
W=\tilde{V}+W^{e}%
\end{equation}
with
\begin{equation}
W^{e}(14,23)=V^{e}(14,23)+i\Sigma^{e}(1,3){\tilde{G}}^{-1}(4,2)+i{\tilde{G}%
}^{-1}(1,3)\Sigma^{e}(4,2)-i\Sigma^{e}(1,3)\Sigma^{e}(4,2).
\end{equation}
The graphical representation of the Eq.~(\ref{bse1}) is shown in
Fig.~\ref{fig2}.

In addition to the static interaction $\tilde{V}$ the effective
interaction $W$ contains diagrams with energy-dependent
self-energies and an energy-dependent \textit{induced
interaction}, where a phonon is exchanged between the particle and
the hole. As discussed in Refs.
~\cite{Tse.89,KTT.97,KST.04,LRT.07} the term
$i\Sigma^{e}(3,1)\Sigma^{e}(2,4)$ has to neglected in the time
blocking approximation if one neglects backward-going propagators
caused by the particle-phonon coupling. This is a reasonable
approximation applied and discussed in many non-relativistic
models (see e.g.
Refs.~\cite{BBBD.79,BB.81,CBG.92,CB.01,SBC.04,Tse.89,KTT.97,KST.04,Tse.05,LT.05}
and references therein). We have to emphasize, however, that all
the RPA ground state correlations are taken into account, because
it is well known that they play a central role for the
conservation of currents and sum rules.

It turns out that both the solution Eq. (\ref{bse1}) $R$ and its
kernel $W$ are singular. Another difficulty arises because
Eq.~(\ref{bse1}) contains integrations over all time points in the
intermediate states. This means that many configurations are
contained in the exact response function which are actually more
complex than $1p1h\otimes phonon$. Tselyaev has introduced in the
Ref.~\cite{Tse.89} the Time Blocking Approximation (TBA), a
special time-projection technique to block the $ph$-propagation
through these complex intermediate states. In this way one obtains
after a Fourier transformation in time a relatively simple
algebraic equation:
\begin{equation}
R(\omega)=\tilde{R}^{0}(\omega)+\tilde{R}^{0}(\omega){\bar
W}(\omega)R(\omega),
\label{respdir}%
\end{equation}
where
\begin{equation}
{\bar W}_{k_{1}k_{4},k_{2}k_{3}}(\omega)=\tilde{V}_{k_{1}k_{4},k_{2}k_{3}}%
+\Phi_{k_{1}k_{4},k_{2}k_{3}}(\omega)-\Phi_{k_{1}k_{4},k_{2}k_{3}}(0)
\label{W-omega}%
\end{equation}
and
\begin{equation}
{\tilde{R}}_{k_{1}k_{4},k_{2}k_{3}}^{0}(\omega)={\tilde{R}}_{k_{1}k_{2}%
}(\omega)\delta_{k_{1}k_{3}}\delta_{k_{2}k_{4}}. \label{R0}%
\end{equation}
${\tilde{R}}_{k_{1}k_{2}}(\omega)$ is the mean field propagator:
\begin{align}
{\tilde{R}}_{ph}(\omega)  &  =-\frac{1}{\varepsilon_{ph}-\omega}%
,\ \ \ \ {\tilde{R}}_{{\alpha}h}(\omega)=-\frac{1}{\varepsilon_{{\alpha}%
h}-\omega},\\
{\tilde{R}}_{hp}(\omega)  &  =-\frac{1}{\varepsilon_{ph}+\omega}%
,\ \ \ \ {\tilde{R}}_{h{\alpha}}(\omega)=-\frac{1}{\varepsilon_{{\alpha}%
h}+\omega},
\end{align}
$\varepsilon_{ph}=\varepsilon_{p}-\varepsilon_{h}$ and $\Phi$ is the
particle-phonon coupling amplitude with the following components:
\begin{align}
\Phi_{ph^{\prime},hp^{\prime}}(\omega)  &  =\sum\limits_{\mu}\Bigl[\delta
_{pp^{\prime}}\sum\limits_{h^{\prime\prime}}\frac{\gamma_{h^{\prime\prime}%
h}^{\mu}\gamma_{h^{\prime\prime}h^{\prime}}^{\mu\ast}}{\omega-\varepsilon
_{p}+\varepsilon_{h^{\prime\prime}}-\Omega^{\mu}}\nonumber\\
&  +\delta_{hh^{\prime}}\Bigl(\sum\limits_{p^{\prime\prime}}\frac
{\gamma_{pp^{\prime\prime}}^{\mu}\gamma_{p^{\prime}p^{\prime\prime}}^{\mu\ast
}}{\omega-\varepsilon_{p^{\prime\prime}}+\varepsilon_{h}-\Omega^{\mu}}%
+\sum\limits_{{\alpha}^{\prime\prime}}\frac{\gamma_{p{\alpha}^{\prime\prime}%
}^{\mu}\gamma_{p^{\prime}{\alpha}^{\prime\prime}}^{\mu\ast}}{\omega
-\varepsilon_{{\alpha}^{\prime\prime}}+\varepsilon_{h}-\Omega^{\mu}%
}\Bigr)\nonumber\\
&  -\Bigl(\frac{\gamma_{pp^{\prime}}^{\mu}\gamma_{hh^{\prime}}^{\mu\ast}%
}{\omega-\varepsilon_{p^{\prime}}+\varepsilon_{h}-\Omega^{\mu}}+\frac
{\gamma_{p^{\prime}p}^{\mu\ast}\gamma_{h^{\prime}h}^{\mu}}{\omega
-\varepsilon_{p}+\varepsilon_{h^{\prime}}-\Omega^{\mu}}\Bigr)\Bigr],
\label{phiph}%
\end{align}%
\begin{align}
\Phi_{\alpha h^{\prime},h\alpha^{\prime}}(\omega)  &  =\sum\limits_{\mu
}\Bigl[\delta_{\alpha\alpha^{\prime}}\sum\limits_{h^{\prime\prime}}%
\frac{\gamma_{h^{\prime\prime}h}^{\mu}\gamma_{h^{\prime\prime}h^{\prime}}%
^{\mu\ast}}{\omega-\varepsilon_{\alpha}+\varepsilon_{h^{\prime\prime}}%
-\Omega^{\mu}}\nonumber\\
&  +\delta_{hh^{\prime}}\Bigl(\sum\limits_{\alpha^{\prime\prime}}\frac
{\gamma_{\alpha\alpha^{\prime\prime}}^{\mu}\gamma_{\alpha^{\prime}%
\alpha^{\prime\prime}}^{\mu\ast}}{\omega-\varepsilon_{\alpha^{\prime\prime}%
}+\varepsilon_{h}-\Omega^{\mu}}+\sum\limits_{p^{\prime\prime}}\frac
{\gamma_{\alpha p^{\prime\prime}}^{\mu}\gamma_{\alpha^{\prime}p^{\prime\prime
}}^{\mu\ast}}{\omega-\varepsilon_{p^{\prime\prime}}+\varepsilon_{h}%
-\Omega^{\mu}}\Bigr)\nonumber\\
&  -\Bigl(\frac{\gamma_{\alpha\alpha^{\prime}}^{\mu}\gamma_{hh^{\prime}}%
^{\mu\ast}}{\omega-\varepsilon_{\alpha^{\prime}}+\varepsilon_{h}-\Omega^{\mu}%
}+\frac{\gamma_{\alpha^{\prime}\alpha}^{\mu\ast}\gamma_{h^{\prime}h}^{\mu}%
}{\omega-\varepsilon_{\alpha}+\varepsilon_{h^{\prime}}-\Omega^{\mu}%
}\Bigr)\Bigr]. \label{phiah}%
\end{align}
As in Fig. \ref{fig1} the indices $p,\alpha$ and $h$ denote the
particles, antiparticles and holes in the Dirac basis. The
amplitudes $\Phi_{ph^{\prime },h{\alpha}}$,
$\Phi_{{\alpha}h^{\prime},hp}$ are neglected, because they have
only are small effect (see Ref.~\cite{LR.06}). The amplitudes
$\Phi _{pp^{\prime},hh^{\prime}}$ and
$\Phi_{hh^{\prime},pp^{\prime}}$ are also disregarded within this
approximation. Therefore, ground state correlations are taken into
account only on the RPA level due to the presence of the $\tilde
{V}_{pp^{\prime},hh^{\prime}}$,
$\tilde{V}_{hh^{\prime},pp^{\prime}}$ terms of the static
interaction in the Eq.~(\ref{respdir}). By definition, the
response function $R(\omega)$ in Eq.~(\ref{respdir}) contains only
configurations which are not more complex than $1p1h\otimes
phonon$.

In Eq.~(\ref{respdir}) $\Phi(0)$ is subtracted from $\Phi(\omega)$. This
corresponds to the subtraction procedure developed by Tselyaev in
Ref.~\cite{Tse.05}. It considers the fact that the effective interaction
$\tilde{V}$ \ being adjusted to experimental data of the ground state contains
effectively many correlations and, in particular, also admixtures of phonons
at the energy $\omega=0$. In the present method, all correlations entering
through the admixture of phonons are taken care of by the additional
interaction term$\ \Phi(\omega)$. To avoid double counting in the effective
interaction the part $\Phi(0)$ is therefore subtracted. This means only the
energy dependence of the phonon coupling is effectively taken into account.

To describe the observed spectrum of the excited nucleus in a weak external
field $D$, as for instance a dipole field, one needs to calculate the strength
function:
\begin{equation}
S(E)=-\frac{1}{\pi}\lim\limits_{\Delta\rightarrow+0}Im\ \Pi_{DD}(E+i\Delta),
\label{strf}%
\end{equation}
expressed through the polarizability
\begin{equation}
\Pi_{DD}(\omega)=D^{\dag}R(\omega)D \label{polarization}%
\end{equation}
A finite imaginary part $\Delta$ of the energy variable is
introduced in the calculations for convenience in order to obtain
a more smoothed envelope of the spectrum. This parameter has the
meaning of an additional artificial width for each excitation.
This width emulates effectively contributions from configurations
which are not taken into account explicitly in this approach.

In order to calculate the strength function it is convenient to convolute
Eq.~(\ref{respdir}) with an external field operator and introduce the
transition density matrix $\delta\rho$ in the external field $D$:
\begin{align}
\delta\rho_{k_{1}k_{2}}(\omega)  &  =\sum\limits_{k_{3}k_{4}}R_{k_{1}%
k_{4},k_{2}k_{3}}(\omega)D_{k_{3}k_{4}},\label{DrhoP}\\
{\delta\rho}_{k_{1}k_{2}}^{0}(\omega)  &  =\sum\limits_{k_{3}k_{4}}{\tilde{R}%
}_{k_{1}k_{4},k_{2}k_{3}}^{0}(\omega)D_{k_{3}k_{4}}, \label{DrhoP0}%
\end{align}
Using Eq.~(\ref{respdir}) we find that $\delta\rho(\omega)$ obeys the
equation
\begin{equation}
\delta\rho(\omega)={\delta\rho}^{0}(\omega)+\tilde{R}^{0}(\omega
)\Bigl(\tilde{V}+\Phi(\omega)-\Phi(0)\Bigr){\delta\rho}(\omega), \label{drho1}%
\end{equation}
and the strength function is expressed as%
\begin{equation}
S(E)=-\frac{1}{\pi}\lim\limits_{\Delta\rightarrow+0}Im\,\text{Tr[}D^{\dag
}\delta\rho(E+i\Delta)\text{]}. \label{strf1}%
\end{equation}

\section{APPLICATIONS}

For the following applications discussed in this section the parameter set
NL3~\cite{NL3} is used for the covariant energy functional. For superfluid
nuclei we use in the pairing channel a simple monopole force with the strength
parameters $G_{\tau}$ ($\tau=p,n$) adjusted to experimental gap parameters for
protons and neutrons. The cut-off energy in the pairing channel is $20$ MeV
both for protons as well as for neutrons. The parameter set NL3 has been
adjusted to ground state properties of a few spherical nuclei more than ten
years ago. In numerous applications it has been shown that it provides on the
mean field level a very good description of ground states and excited states
of nuclei all over the periodic table~\cite{LRR.99}. In a recent
investigation~\cite{LKF.09} its parameters have been slightly modified and
several small deficiencies have been eliminated.

\subsection{Single-particle spectra in the Pb-region}

\label{appl-1}

\begin{table}[ptb]
\caption{Energies $\varepsilon_{k}^{(d)}$ and spectroscopic factors
$S_{k}^{(d)}$ of the dominant neutron levels in $^{208}$Pb calculated
in the strongly restricted particle-phonon space. $ph\alpha$ denotes
full the calculation, $p\alpha$ ($h$) is the version without
backwards going terms, and $ph$ is the version without contribution
of the antiparticle states in the
self-energy (see text for details).}%
\label{table1}
\begin{center}
\vspace{3mm} \tabcolsep=1.25em \renewcommand{\arraystretch}{1.1}%
\begin{tabular}
[c]{cccccccc}\hline\hline
State $k$ & $\varepsilon_{k}$, MeV & \multicolumn{3}{c}{$\varepsilon_{k}%
^{(d)}$, MeV} & \multicolumn{3}{c}{$S_{k}^{(d)}$}\\\hline Particle &
\  & $ph\alpha$ & $p\alpha$ & $ph$ & $ph\alpha$ & $p\alpha$ &
$ph$\\\hline
2g9/2 & -2.50 & -2.85 & -3.14 & -2.88 & 0.89 & 0.92 & 0.89\\
1i11/2 & -2.97 & -2.82 & -3.20 & -2.90 & 0.94 & 0.97 & 0.94\\
1j15/2 & -0.48 & -1.16 & -1.33 & -1.21 & 0.70 & 0.74 & 0.70\\
3d5/2 & -0.63 & -0.96 & -1.05 & -0.98 & 0.93 & 0.94 & 0.93\\
4s1/2 & -0.36 & -0.88 & -0.92 & -0.89 & 0.93 & 0.93 & 0.93\\
2g7/2 & -0.56 & -0.71 & -0.90 & -0.76 & 0.92 & 0.94 & 0.92\\
3d3/2 & -0.02 & -0.35 & -0.42 & -0.37 & 0.93 & 0.93 & 0.93\\\hline
Hole & \  & $ph\alpha$ & $h$ & $ph$ & $ph\alpha$ & $h$ & $ph$\\\hline
3p1/2 & -7.66 & -7.67 & -7.40 & -7.70 & 0.96 & 0.98 & 0.96\\
2f5/2 & -9.09 & -8.97 & -8.71 & -9.02 & 0.93 & 0.96 & 0.93\\
3p3/2 & -8.40 & -8.20 & -7.87 & -8.22 & 0.90 & 0.94 & 0.90\\
1i13/2 & -9.59 & -9.30 & -9.07 & -9.36 & 0.90 & 0.92 & 0.89\\
2f7/2 & -11.11 & -10.20 & -9.98 & -10.22 & 0.72 & 0.76 & 0.72\\
(1h9/2)$_{1}$ & -13.38 & -13.32 & -13.23 & -13.34 & 0.52 & 0.47 & 0.53\\
(1h9/2)$_{2}$ & \  & -12.48 & -12.42 & -12.49 & 0.31 & 0.39 &
0.29\\\hline\hline
\end{tabular}
\end{center}
\end{table}

In this section we discuss the changes of the single-particle
spectra of the odd mass nuclei $^{207}$Pb, $^{209}$Pb, $^{207}$Tl
and $^{209}$Bi if the coupling to low lying collective vibrations
of the surface is taken into account. In order to keep the
numerical effort in reasonable limits in a first investigation
only the most collective phonons with spin and parity $J^{\pi
}=2^{+},$ $3^{-},$ $4^{+},$ $5^{-},$ $6^{+}$ below the neutron
separation energy and a reduced number of single-particle states
with positive energy (particles or holes) is taken into account in
the solution of the Dyson equation (\ref{Dyson}). This reduces
strongly the number of poles in the self-energy of Eq.
(\ref{mo1}). The numerical results obtained in these
investigations are compiled in the Table \ref{table1}. For the
first shell of neutron levels above ('particle') and below
('hole') the Fermi level three versions are given: in the version
$ph\alpha$ the index $n$ in Eq. (\ref{mo1}) includes all
contributions from intermediate states above the Fermi level $p$,
below the Fermi level $h$ and in the Dirac sea $\alpha$. Version
$p\alpha$ (for particles) or $h$ (for holes) excludes the backward
going diagrams, and the third version $ph$ does not contain
antiparticle intermediate states in (\ref{mo1}). In this way, one
can see that the effects of ground state correlations (GSC) caused
by the particle-phonon coupling and neglected in the second
version are significant and it is essential to take them into
account in a realistic calculation. On the other hand, the
contribution of the antiparticle subspace to the self-energy is
quantitatively not of great importance. This can be understood by
the large values of the energy denominators in Eq. (\ref{mo1}) for
these configurations. Thus it is justified to disregard them in
the following calculations. Notice, however, that version $ph$
does not eliminate the effects of the Dirac sea completely since
the phonon vertices still contain this contribution. As it has
been discussed in Ref.~\cite{RMG.01} these terms play an important
role in a proper treatment of relativistic RPA. Otherwise it is
not possible to obtain reasonable properties for the isoscalar
modes within RRPA.

Next we show results where the contribution of the antiparticle
subspace to the self-energy are neglected. In this case one is
able to enlarge the particle-hole basis considerably by taking
into account particle-hole configurations far away from the Fermi
surface. This increases the collectivity of the phonons and,
consequently, the strength of the particle-vibrational coupling.
The phonon basis was also enriched by including higher-lying modes
up to 35 MeV. Solving the Dyson equation one finds a fragmentation
of the single-particle states and a corresponding reduction of the
single-particle strength. For the levels one major shell below and
one shell above the Fermi surface one finds always one dominant
level, which is shifted against the corresponding single-particle
energy without particle-phonon coupling. Almost all the levels are
moving downwards providing thus a considerably better agreement
with experimental energies then the pure RMF states. In the next
shells further away from the Fermi surface almost all the
single-particle levels turn out to be strongly fragmented due to
phonon coupling and it is no longer possible to determine the
dominant levels in these shells, in other words, the concept of
Landau quasi-particles is defined only in the neighborhood of the
Fermi level and it breaks down at larger distances.

\begin{figure}[ptb]
\begin{center}
\includegraphics*[scale=0.8]{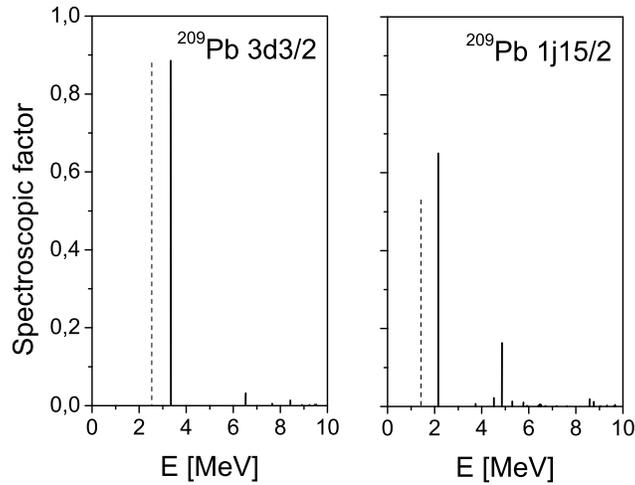}
\end{center}
\vspace{-1cm}%
\caption{Single-particle strength distribution for the $3d_{3/2}$
(left panel) and $1j_{15/2}$ (right panel) states in $^{209}$Pb
obtained in the calculations (solid lines) and the experimental
strengths of the respective
dominant levels (dashed lines).}%
\label{fig3}%
\end{figure}

In Fig.~\ref{fig3} we show as an example the 3d$_{3/2}$ and the
1j$_{15/2}$ levels in the nucleus $^{209}$Pb. In both cases the
single-particle strength is distributed over about two thousand
states but most of them are vanishingly small. Thus only the
states with the strength exceeding 10$^{-3}$ are drown. The state
3d$_{3/2}$ has a pronounced single-particle structure with a
single-particle strength close to 0.9. On the other side the
1j$_{15/2}$ is more fragmented. The experimental strength of the
dominant levels are shown with dashed lines.

\begin{figure}[ptb]
\begin{center}
\includegraphics*[scale=1.0]{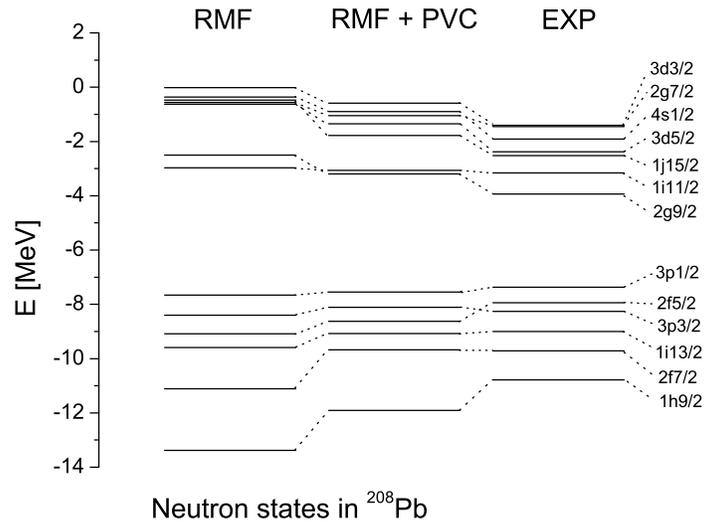}
\end{center}
\vspace{-1cm}%
\caption{Neutron single-particle states in Pb$^{208}$: the pure RMF
spectrum (left column), the levels computed within RMF with allowance
for the
particle-vibration coupling (center) and the experimental spectrum (right).}%
\label{fig4}%
\end{figure}

To illustrate the shifts in the level schemes of the dominant poles as
compared to the RMF results we show as an example in Figs.~\ref{fig4} the
single-particle spectrum for neutrons. The spectrum calculated with the
energy-dependent correction (RMF+PVC) demonstrates a pronounced increase of
the level density around the Fermi surface of $^{208}$Pb compared the pure RMF
spectra. In some cases the order of levels is inverted and the observed
sequence is reproduced as for instance for the$\ 1j_{15/2}$ and the $3d_{5/2}$
neutron states. Another and more important example is the inversion of the
$2g_{9/2}$ and $1i_{11/2}$ neutron states which reproduces the spin of the
$^{209}$Pb ground state.

In order to quantify these results we calculate the average
distance between two levels in the spectrum shown in
Fig.~\ref{fig4}. One finds for the neutrons 1.0 (RMF), \ 0.83
(RMF+PCV) and 0.76 (EXP) in units of MeV. This corresponds to a
level density of 1.0 (RMF), 1.20 (RMF+PCV) and 1.31 (EXP) in units
of MeV$^{-1}$. The level density in the neighborhood of the Fermi
surface is therefore in RMF-calculations by a factor 0.76 smaller
than the experimental value. Taking into account
particle-vibrational coupling we find only a reduction of \ 0.92.
Assuming an effective mass close to 1 for the experiment, and
taking into account that the level density at the Fermi surface is
proportional to $m^{\ast}/m$, this corresponds to an effective
mass $m^{\ast}/m\approx0.76$ for the RMF and
$m^{\ast}/m\approx0.92$ for the RMF+PCV calculations. For the
protons the situation is similar.

Jaminon and Mahaux~ have discussed in Refs. \cite{JM.89,JM.90} the concept of
the effective mass in the case of RMF theory. On one side one has the well
known Dirac mass%
\begin{equation}
m_{D}=m+\tilde{\Sigma}_{s}(\mathbf{r),}%
\end{equation}
which is determined by the scalar field $\tilde{\Sigma}_{s}$.
Since we do not use an isovector scalar field for the present
parameter set NL3 the Dirac mass is in these calculations
identical for protons and neutrons. However, this quantity should
not be compared with the effective mass determined empirically
from a non-relativistic analysis of scattering data and of bound
states. From a non-relativistic approximation of the Dirac
equations one finds that the mass
\begin{equation}
m_{eff}=m-\tilde{\Sigma}_{0}%
\end{equation}
should be used for this purpose. Here $\tilde{\Sigma}_{0}$ is the time-like
component of the Lorentz vector field determined by the exchange of $\omega$-
and $\rho$-mesons.

In symmetric nuclear matter we find for NL3: $m_{D}/m=0.60$ and $m_{eff}%
/m=0.67$. The latter value is smaller then the values $m^{\ast}/m\approx0.71$
for protons and $m^{\ast}/m\approx0.76$ for neutrons deduced from the
calculated spectrum around the Fermi surface in simple RMF theory. Following
similar arguments we would obtain for RMF+PVC calculations an average
effective mass of $0.89$. This is obviously still too low as compared to the
experimental value.

On the other hand, around the Fermi surface where relativistic kinematic
effects are not significant the RMF+PVC spectrum can be characterized by the
effective mass deduced from the Schr\"{o}dinger equation which is a
non-relativistic limit of the Dirac equation (\ref{Deq1}). In this
approximation one can calculate the state-dependent E-mass ${\bar{m}}/m^{RMF}$
which is the inverted spectroscopic factor of the dominant level $\lambda$:%
\begin{equation}
\frac{{\bar{m}}_{k}}{m^{RMF}}=\bigl [S_{k}^{(\lambda)}\bigr]^{-1}.
\end{equation}
For the calculated RMF+PVC spectrum the averaged E-masses are 1.26
for neutrons and 1.41 for protons if one takes into account all
the states with spectroscopic factors larger then 0.5, i. e. good
single-particle states. Thus, the energy dependence of the
self-energy increases the RMF neutron and proton effective masses
up to the values 0.96 and 1.0, respectively.

Although the problem of particle-vibration coupling in nuclei has
a long history and it was considered in a number of works, most of
them are based on a non-relativistic treatment of the nuclear
many-body problem. Only in a relatively recent investigation in
Ref.~\cite{VNR.02} a correction of the RMF single-particle
spectrum was undertaken in a phenomenological way assuming a
linear dependence of the self-energy near the Fermi surface. The
corresponding coupling constants were determined by a fit to
nuclear ground state properties. Despite the fact that the present
approach is fully microscopic without any additional parameter
adjusted to experiment it shows good agreement with the results of
Ref.~\cite{VNR.02} for the spectrum of $^{208}$Pb. The shift
caused by the phenomenological particle-vibrational coupling in
Ref.~\cite{VNR.02} is only slightly larger than in the present
investigation.

Non-relativistic microscopic investigations of
particle-vibrational coupling can be divided into two major
groups. The first group~\cite{RW.73,KT.86,Pla.81,HS.76} uses a
phenomenological single-particle input to reproduce the
experimental spectrum and has therefore to exclude the
contribution of the particle-vibration coupling from the full
self-energy to find the 'bare' spectrum. Usually these older
approaches take into consideration only a relatively small number
of collective low-lying phonons and use a particle-vibration
coupling model~\cite{BM.75}. This restriction to only low-lying
modes produces shifts less then 1 MeV. However, as it was shown in
Ref.~\cite{HS.76}, enlarging of the phonon space with high-lying
vibrations leads to very strong shifts of the single-particle
levels up to 4 MeV, and no saturation is observed with respect to
the dimension of the phonon space.

The second group of approaches (see, for instance
Refs.~\cite{PRS.80,BG.80}) starts from a \ self-consistent
Hartree-Fock description and applies perturbation theory to
calculate the particle-vibration contribution to the full
self-energy. In such self-consistent methods it is more justified
to enlarge the phonon space. It was shown, for instance,
in~\cite{BG.80} that the contribution of the isovector modes is
noticeably smaller than the isoscalar ones. The detailed
investigation of the relative importance of the high multipole
states was performed in~\cite{PRS.80}. Because of the larger
phonon space the typical shifts of the single-particle levels in
$^{208}$Pb are about 1-2 MeV.

As for the spectroscopic factors, all the approaches predict similar values
because these factors are not very sensitive to the details of the calculation schemes.

\subsection{The strength functions of collective excitations in closed shell
nuclei}

The solution of the Bethe-Salpeter equation (\ref{respdir}) allows
to calculate the nuclear response to external multipole fields and
the strength functions of the corresponding collective
excitations. As in the last section we show applications based on
the density functional NL3 with a monopole force in the pairing
channel. A small artificial width of 200 keV is introduced as an
imaginary part of the energy variable $\omega$ to have a smooth
envelope of the calculated curves. The energies and amplitudes of
the most collective phonon modes with spin and parity 2$^{+}$,
3$^{-}$, 4$^{+}$, 5$^{-}$, 6$^{+}$ are calculated with the same
restrictions and selected using the same criterion as in the last
section and in many other non-relativistic investigations in this
context. Only the phonons with energies below the neutron
separation energy enter the phonon space since the contributions
of the higher-lying modes are found to be small.

On all three stages of these calculations the same energy functional, i.e. the
same relativistic nucleon-nucleon interaction $\tilde{V}$ \ (\ref{Vstatic})
has been employed. The vertices $\gamma_{k_{1}k_{2}}^{\mu}$ (\ref{phonon})
entering the term $\Phi(\omega)$ in Eq. (\ref{phiph}) are calculated with the
same force. Therefore no further parameters are needed. The scheme is fully consistent.

The subtraction procedure developed by Tselyaev in the
Ref.~\cite{Tse.05} for the self-consistent scheme removes the static
contribution of the particle-phonon coupling from the
$ph$-interaction. It takes into account only the additional energy
dependence introduced by the dynamics of the system. It has been
found in the calculations of Refs.~\cite{LT.05} as well as in the
calculations of the Ref.~\cite{LRT.08} that within the relatively
large energy interval (0 - 30 MeV) the subtraction procedure provides
a rather small increase of the mean energy of the giant dipole
resonance (0.8 MeV for lead region) and gives rise to the change by a
few percents in the sum rule. This procedure restores the response at
zero energy and therefore it does not disturb the symmetry properties
of the RRPA calculations. The zero energy modes connected with the
spontaneous symmetry breaking in the mean field solutions, as for
instance the translational mode in the dipole case, remain at exactly
the same position after the inclusion of the particle-vibration
coupling. In practice, however, because of the limited number of
oscillator shells in the calculations this state is found already in
RRPA without particle-vibration coupling at a few hundreds keV above
zero. In cases, where the results depend strongly on a proper
separation of this spurious state, as for instance for investigations
of the pygmy dipole resonance in neutron rich systems
\cite{LRV.07,LRT.09} one has to include a large number of
$ph$-configurations in the RRPA solution.

\begin{figure}[ptb]
\begin{center}
\includegraphics*[scale=1.0]{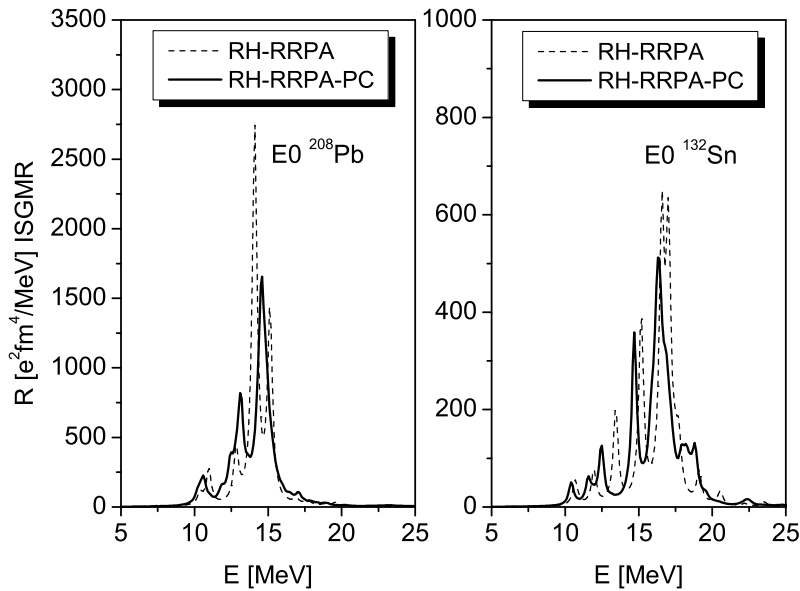}
\end{center}
\vspace{-1cm}%
\caption{Isoscalar monopole resonance in $^{208}$Pb and $^{132}$Sn
obtained within two approaches: RRPA (dashed line) and RRPA with
particle-phonon coupling RRPA-PC (solid line). Both computations have
been performed with relativistic Hartree (RH) mean field and employ
NL3 parameter set for RMF
forces.}%
\label{fig5}%
\end{figure}

\begin{table}[ptb]
\caption{Lorentz fit parameters of isoscalar E0 strength function in $^{208}%
$Pb and $^{132}$Sn calculated within RRPA and RRPA extended by the
particle-phonon coupling model (RRPA-PC) as compared to experimental
data. The
fit has been carried out in the interval from $B_{n}$ to roughly 20 MeV}.%
\label{table2}
\par
\begin{center}
\vspace{3mm} \tabcolsep=2.15em \renewcommand{\arraystretch}{1.1}%
\begin{tabular}
[c]{cccc}\hline\hline &  & $<$E$>$ (MeV) & $\Gamma$ (MeV)\\\hline
& RRPA & 14.16 & 1.71\\
$^{208}$Pb & RRPA-PC & 14.05 & 2.36\\
& Exp.~\cite{SY.93} & 13.73(20) & 2.58(20)\\\hline
& RRPA & 16.10 & 2.63\\
$^{132}$Sn & RRPA-PC & 16.01 & 3.09\\\hline\hline
\end{tabular}
\end{center}
\end{table}

In Fig.~\ref{fig5} we show the calculated strength functions for the isoscalar
monopole resonance in $^{208}$Pb and $^{132}$Sn. The fragmentation of the
resonance caused by the particle-phonon coupling is clearly demonstrated
although the spreading width of the monopole resonance is not large because of
a strong cancellation between the self-energy diagrams and diagrams with the
phonon exchange (see Fig.~\ref{fig2}). This fact has also been discussed in
detail in Refs.~\cite{BBBD.79,BB.81} and it is not disturbed by the
subtraction procedure because this cancellation takes place as well in
$\Phi(\omega)$ as in $\Phi(0)$.

In order to compare the spreading of the theoretical strength
distributions with experimental data we show in Table~\ref{table2}
mean energies $\langle E\rangle$ and widths parameters $\Gamma$
obtained by fitting the theoretical strength distribution in a
certain energy interval to a Lorentz curve in the same way as it
has been done in the experimental investigations. The experimental
values shown in Table \ref{table2} are derived in the
Ref.~\cite{SY.93} from the evaluation of a series of data obtained
in different experiments for the isoscalar monopole resonance in
$^{208}$Pb.

\begin{figure}[t]
\begin{center}
\includegraphics*[scale=1.0]{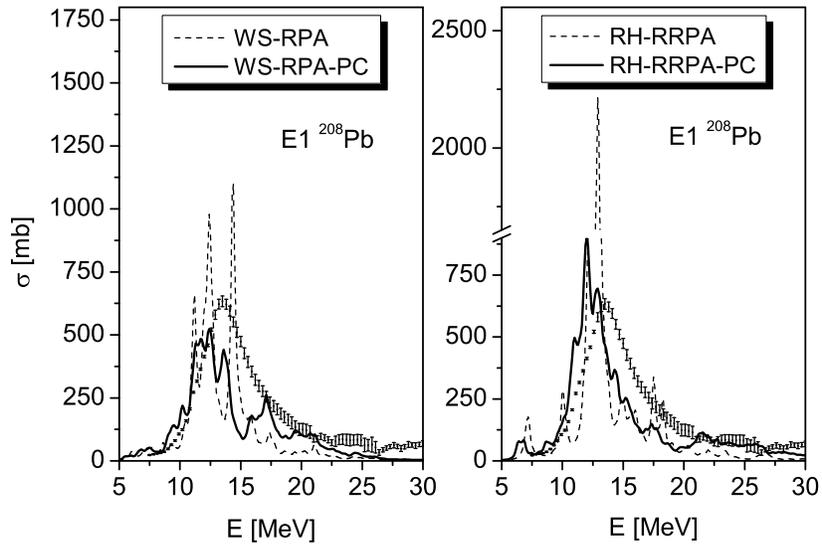}
\end{center}
\vspace{-1cm}%
\caption{The isovector E1 resonance in $^{208}$Pb. Details are given
in the
text.}%
\label{fig6}%
\end{figure}

\begin{figure}[t]
\begin{center}
\includegraphics*[scale=1.0]{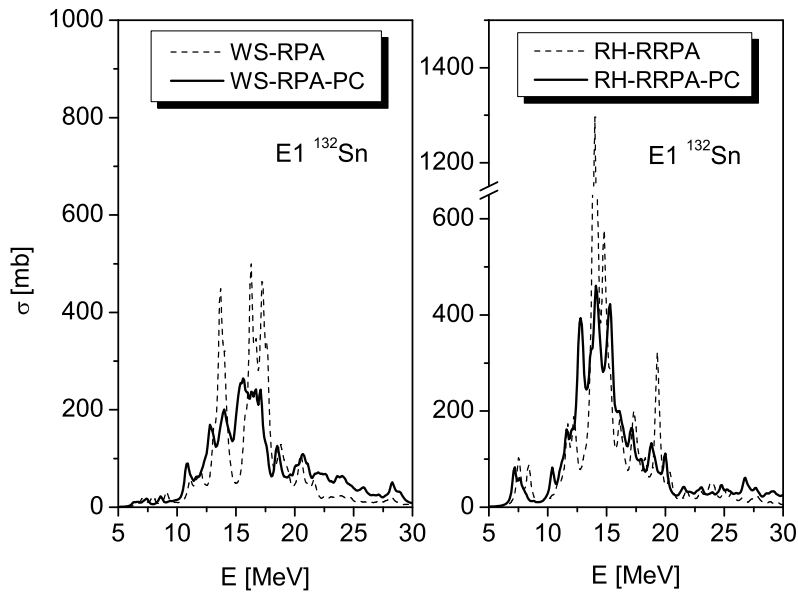}
\end{center}
\vspace{-1cm}%
\caption{The same as in Fig.~\ref{fig6} but for $^{132}$Sn.}%
\label{fig7}%
\end{figure}

Figs.~\ref{fig6} and \ref{fig7} present calculated photoabsorption cross
sections
\begin{equation}
\sigma_{E1}(E)={\frac{{16\pi^{3}e^{2}}}{{9\hbar c}}}E~S_{E1}(E)
\label{cs}
\end{equation}
for the isovector dipole resonance in $^{208}$Pb and in
$^{132}$Sn. The left panels give the results obtained within the
non-relativistic approach with a Woods-Saxon (WS) single-particle
potential and Landau-Migdal (LM) forces described in
Ref.~\cite{LT.05}. They are compared in the right panel with the
relativistic fully consistent theory of Ref.~\cite{LRT.07}. (R)RPA
calculations are shown by the dashed curves, (R)RPA extended by
the phonon coupling ((R)RPA-PC) calculations -- by the thick solid
curves. In Fig. \ref{fig6} we have also displayed experimental
data with error bars taken from Ref.~\cite{ripl}. In both
calculations, relativistic (right panel) and non-relativistic
(left panel), the continuum is taken into account only in a
discrete approximation, which is very reliable for heavy nuclei.
As discussed in Ref. \cite{GRT.90}, the Dirac equation
(\ref{dirac-basis}) is solved in an oscillator basis. To make the
comparison reasonable calculations within the non-relativistic
framework have been performed with box boundary conditions for the
Schr\"{o}dinger equation in $r$-space which ensures completeness
of the single-particle basis.

The corresponding Lorentz fit parameters in the two energy intervals:
$B_{n}-25$ MeV and $0-30$ MeV ($B_{n}$ is the neutron separation
energy) are included in Table~\ref{table3} and they are compared with
the data of Ref.~\cite{Adr.05,ripl}. We notice that the inclusion of
particle-phonon coupling in the RRPA calculation induces a pronounced
fragmentation of the photoabsorption cross sections, and brings the
width of the GDR in much better agreement with the data, both for
$^{208}$Pb and $^{132}$Sn.

The fragmentation of the resonance introduced by the particle-phonon coupling
is clearly demonstrated in both cases. Also, one finds more or less the same
level of agreement between theory and experimental data for these two
calculations. In the case of the isovector E1 resonance in $^{132}$Sn this is,
however, not so clear because the cross section and the integral
characteristics of the resonance obtained in the experiment of
Ref.~\cite{Adr.05} are given with relatively large error bars. In $^{208}$Pb
the self-consistent relativistic approach reproduces the shape of the giant
dipole resonance much better than the non-relativistic one although the whole
resonance is about 0.5 MeV shifted to lower energies with respect to the
experiment. As one can see from the Fig.~\ref{fig6} and Table \ref{table3}, we
observe some shift already in the RRPA calculation, which is determined by the
properties of the NL3 forces. Improvement of the forces, for instance, the use
of the density dependent versions~\cite{DD-ME1,DD-ME2} of the RMF should bring
the E1 mean energy in better agreement with the data.

\begin{table}[ptb]
\caption{Lorentz fit parameters in the two energy intervals:
$B_{n}-25$ MeV and $0-30$ MeV, for the E1 photo absorption cross
sections in $^{208}$Pb and $^{132}$Sn, calculated with the RRPA, and
with the RRPA extended to include
the particle-phonon coupling (RRPA-PC), compared to data.}%
\label{table3}
\begin{center}
\tabcolsep=1.1em \renewcommand{\arraystretch}{1.1}%
\begin{tabular}
[c]{cccccccc}\hline\hline
&  & \multicolumn{3}{c}{$B_{n}$ - 25 MeV} & \multicolumn{3}{c}{0 - 30 MeV}\\
&  & $<$E$>$ & $\Gamma$ & EWSR & $<$E$>$ & $\Gamma$ & EWSR\\
&  & (MeV) & (MeV) & (\%) & (MeV) & (MeV) & (\%)\\\hline
& RRPA & 13.1 & 2.4 & 121 & 12.9 & 2.0 & 128\\
$^{208}$Pb & RRPA-PC & 12.9 & 4.3 & 119 & 13.2 & 3.0 & 128\\
& Exp.~\cite{ripl} & 13.4 & 4.1 & 117 &  &  & 125(8)\\\hline
& RRPA & 14.7 & 3.3 & 116 & 14.5 & 2.6 & 126\\
$^{132}$Sn & RRPA-PC & 14.4 & 4.0 & 112 & 14.6 & 3.2 & 126\\
& Exp.~\cite{Adr.05} & 16.1(7) & 4.7(2.1) & 125(32) &  &  &
\\\hline\hline
\end{tabular}
\end{center}
\end{table}

However, there is an essential difference between the fully
self-consistent relativistic calculations and the non-relativistic
approach: in non-relativistic approach discussed in Ref.~\cite{LT.05}
one introduces on all three stages of the calculation
phenomenological parameters, which have to be adjusted to
experimental data: first, the Woods-Saxon parameters as, for
instance, the well depth are varied to obtain single-particle levels
close to the experimental values, second, one of the parameters of
the Landau-Migdal force is adjusted to get phonon energies at the
experimental positions (for each mode) and, third, another
Landau-Migdal force parameter is varied to reproduce the centroid of
the giant resonance. Although the varying of the parameters is
performed in relatively narrow limits, it is necessary to obtain
realistic results. In contrast, in the relativistic fully consistent
approach no adjustment of additional parameters is necessary. Of
course, the underlying energy functional has been determined in a
phenomenological way by a fit to experimental ground state properties
of characteristic nuclei. However, it is of universal nature and the
same parameters are used for investigations of many nuclear
properties all over the periodic table. The predictive power of this
scheme is therefore much higher than that of the present
semi-phenomenological approach discussed, for instance, in
Ref.~\cite{LT.05}.

\subsection{Collective excitations in systems with pairing}

Pairing correlations play an essential role in all open shell
nuclei and apart from the vicinity of the very few doubly magic
configurations nuclei show superfluidity all over the periodic
table. In a theoretical description this fact can be taken into
account by Bogoliubov's quasiparticles. Many-body theories for
normal systems are thus relatively easily extended to the case of
superfluid nuclei. Combining creation and annihilation operators
$a^{+}$ and $a$ to a two-component operator, operators of the type
$a^{+}a$, $a^{+}a^{+}$, $aa$ are replaced by super-matrices of
rank 2 and the form of the equations stays nearly unchanged. This
very elegant method has been introduced already half a century ago
in Ref. \cite{Gor.58} and over the years it has been used for
various many-body approximation schemes in non-relativistic
systems as for instance in Refs. \cite{Val.61,BW.63,RRE.84,BR.86}.
Response theory with the Time Blocking Approximation (TBA)
introduced for normal non-relativistic systems in Ref.
\cite{Tse.89} has been extended in Refs. \cite{Tse.05,LT.05} to
QTBA for superfluid systems and in Ref. \cite{LRT.08} to RQTBA for
relativistic superfluid systems.

In the following we discuss several applications of RQRPA and of RQTBA in the
chain of spherical even-even semi-magic nuclei with $Z=50$. We show
calculations of the isovector dipole spectrum in the giant dipole resonance
region and in the low-lying energy region in the two approximations.

As discussed before the effective interaction
$\Phi(\omega)$-$\Phi(0)$ takes into account only the additional
energy dependence introduced by the dynamics of the system. It has
been found in relativistic~\cite{LRT.08} as well as in
non-relativistic calculations~\cite{LT.05} that within a
relatively large energy interval (0 - 30 MeV) the subtraction
procedure provides a rather small but noticeable increase of the
mean energy of the giant dipole resonance (about 0.7 MeV for tin
region) and gives rise to changes by a few percents in the sum
rule. The absolute value of the energy shift produced by the
subtraction of $\Phi(0)$ in Eq.~(\ref{W-omega}) is comparable with
but not exactly equal to the absolute value of the shift produced
by the dynamical part of the interaction amplitude $\Phi(\omega)$
which always reduces the mean energy of the resonance. The
subtraction procedure restores the response at zero energy and,
therefore, it does not disturb the symmetry properties of the
RQRPA calculations. The zero energy modes connected with the
spontaneous symmetry breaking in the mean field solutions, as, for
instance, the translational mode in the dipole case, remain at
exactly the same positions after the inclusion of the
quasiparticle-vibration coupling.

\begin{figure}[ptb]
\begin{center}
\includegraphics*[scale=1.5]{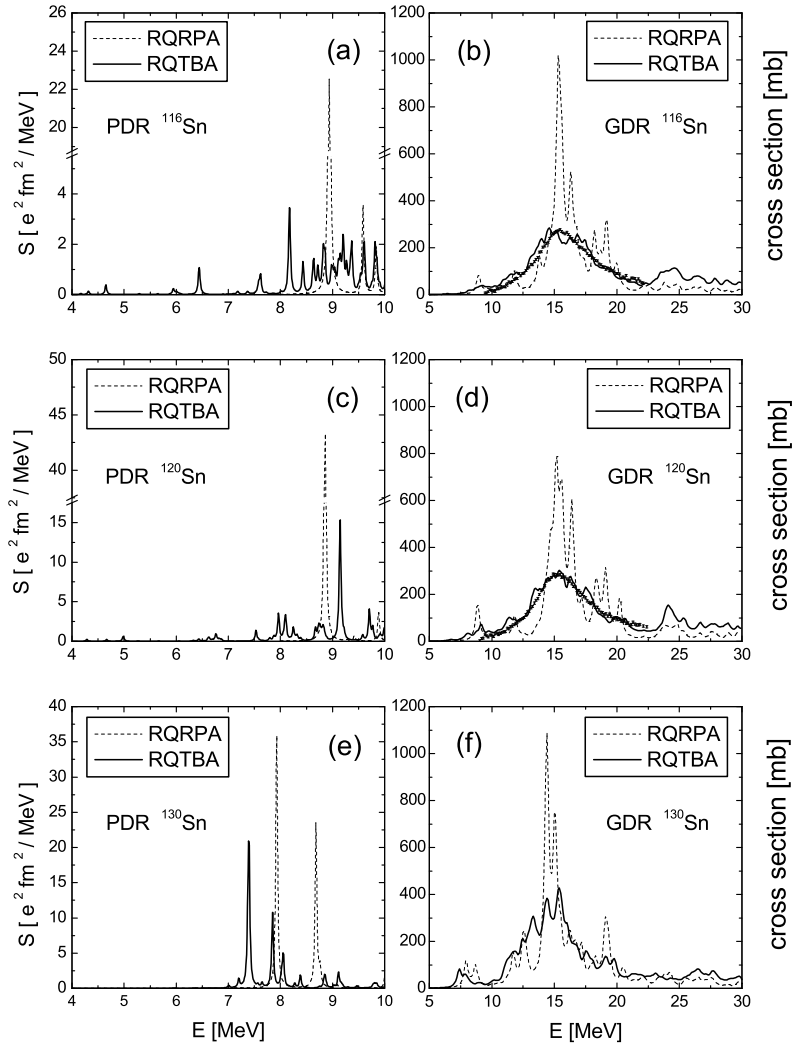}
\end{center}
\vspace{-1cm}%
\caption{The calculated dipole spectra for the heavier tin isotopes
$^{116}$Sn, $^{120}$Sn, $^{130}$Sn, compared to data of
Ref.~\cite{exfor} for $^{116,120}$Sn. Right panels (b, d, f): photo
absorption cross sections computed with the artificial width 200 keV.
Left panels (a, c, e): the low-lying portions of the corresponding
spectra in terms of the strength function, calculated with 20 keV
smearing. Calculations within the RQRPA are
shown by the dashed curves, and the RQTBA - by the solid curves.}%
\label{fig8}%
\end{figure}

In Fig.~\ref{fig8} we show dipole spectra for the tin isotopes
$^{116}$Sn, $^{120}$Sn, $^{130}$Sn. The right panels show the
photo absorption cross section (\ref{cs}) which is determined by
the dipole strength function $S_{E1}$. It is calculated,
analogously to Eq. (\ref{strf1}), with the usual isovector dipole
operator. The left panels show the low-lying parts of the
corresponding spectrum in terms of the strength function. A small
imaginary part of 20 keV is used for the energy variable, in order
to see the fine structure of the spectrum and sometimes individual
levels in this region. RQRPA calculations are shown by dashed
curves and the RQTBA by the solid curves. Experimental data are
taken from the EXFOR database ~\cite{exfor}.

These figures clearly demonstrate how the two-quasiparticle
states, which are responsible for the spectrum of the RQRPA
excitations, are fragmented through the coupling to the collective
vibrational states. The effect of the particle-vibration coupling
on the low-lying dipole strength below and around the neutron
threshold within the presented approach is shown in the left
panels of the Fig.~\ref{fig8}. Such calculations give us an
example how the low-lying strength develops with the increase of
the neutron excess. It is also found that the presence of pairing
correlations causes a noticeably stronger fragmentation of both
the GDR and the PDR modes as compared to the case of a normal
system discussed above. This effect has the two reasons. First,
pairing correlations lead to a diffuseness of the Fermi surface
and, thus, increase the number of possible $2qp\otimes phonon$
configurations, and second, pairing correlations cause a
considerable lowering of the energies and increased transition
probabilities of the lowest 2$^{+}$ states. In spherical
open-shell medium mass nuclei the highly collective first 2$^{+}$
states appear at energies around 1 MeV (and they are usually well
reproduced in RQRPA~\cite{AR.06}) whereas in magic nuclei and
often in nuclei near the shell closures they appear much higher,
at about 3-4 MeV and have considerably reduced transition
probabilities. This causes a strong configuration mixing in the
case of presence of very low-lying vibrational states. These modes
admix to others, in particular, to the GDR and the PDR and the
lower their energies and the higher their transition probabilities
are, the stronger fragmentation they cause.

A systematic analysis of the transition densities of the RQRPA and the RQTBA
states shows that the $2qp$ transition densities in the broad low-lying energy
region dominated by the fragmentation of the RQRPA pygmy mode have a very
similar behavior as the initial RQRPA state: proton and neutron components
oscillate in phase in the nuclear interior and neutron components dominate on
the surface in nuclei with noticeable neutron excess.

\section{CONCLUSIONS}

We have given an overview over recent efforts to combine two
theoretical methods for the description of the quantum-mechanical
many-body problem of nuclear physics, Covariant Density Functional
Theory (CDFT) and Landau-Migdal Theory of Finite Fermi Systems
(TFFS). Both methods are very successful and they are claimed to
provide in principle an exact description. In practice, however,
there are limitations. Both methods use phenomenological input.
DFT can only be applied to physical quantities, which can be
expressed in terms of the single-particle density and in
self-bound systems such as nuclei DFT is based on the intrinsic
density, a concept, which requires additional approximations. In
particular the self-energy used in DFT theory does not depend on
the energy. Landau-Migdal theory on the other side restrains
itself from calculating ground state properties, but it goes far
beyond the mean field approach and takes into account couplings to
complex configurations. There are also similarities for these two
methods. Both are based on a single-particle description, i.e. on
the motion of independent particles. In density functional theory
an average field in introduced as a vehicle in order to take into
account shell effects. However, the single-particle energies
themselves are not observables in the strict sense. Landau-Migdal
theory uses quasiparticles as the exact eigenstates of the
$A\pm1$-systems. In both cases the self energies are given as the
first derivatives of the total energy with respect to the density
and the effective interactions between the particles are the
second derivatives of this quantity.

The combination of CDFT and TFFS described in this manuscript
starts from the covariant density functional. No further
parameters are needed. This functional is used to describe the
ground state properties and the parameters of the functional are
adjusted to experimental data of ground states of several nuclei.
On this level the self-energy does not depend on the energy and
Landau-Migdal theory is used to introduce an energy dependence
with a particle-vibrational scheme. The properties of the phonons
needed for the calculation of the energy-dependent part of the
self-energy are phonon energies and phonon-nucleon vertices. They
are calculated with the static effective interaction obtained as
the second derivative of the density functional. No additional
parameters are needed. The essential equation of the Landau-Migdal
theory is the response equation. The effective interaction to
calculate the full response is the derivative of the self-energy
with respect to the density and this means that one obtains in
addition to the static interaction resulting from the energy-%
independent part an induced interaction resulting from the energy-%
dependent part of the self-energy. Of course there would be double
counting, because many of the correlations induced by the coupling
of virtual phonons have also contributions at the ground state
energy. Therefore a subtraction method is introduced, which
removes from the induced interaction at finite energy its value at
zero energy, i.e. after this subtraction, the induced interaction
vanishes at the ground state and takes into account only its
energy dependence. Therefore this subtraction procedure guarantees
that one does not need to readjust the parameters of the density
functional, because the effects of particle-vibrational coupling
vanish at the ground state of the even-even system, where the
parameters are adjusted.

We have discussed several applications of this method, as the
fragmentation of single-particle energies in odd-mass nuclei in
the vicinity of a double magic configuration. Close to the Fermi
surface there is always a dominant pole with a reduced single-%
particle strength and many other poles with rather small strength.
The dominant pole is shifted in the direction of the Fermi
surface, i.e. the level density at the Fermi surface is increased.
The effective mass derived from this level density in the Pb
region is considerably increased, but there is still room for an
additional energy dependence of the self-energy not taken into
account by the coupling to surface vibrations. In addition we
discussed several solutions of the response equations for nuclei
in an external field. This allows to calculate the strength
functions with respect to an external operator and the photo
absorption cross sections in the correlated system. If one takes
into account only the static part of the interaction one finds the
usual RPA or QRPA results of time-dependent density functional
theory, which reproduces the position of the resonances rather
well, but it cannot account for the width that has its origin in
the coupling to more complicated configurations. The energy-%
dependent part of the interaction includes this coupling and
therefore it induces again a fragmentation of the rather sharp
resonance peaks in RPA of QRPA over many complex configurations.
Because of the subtraction procedure the position of the
resonances is not changed very much, but the width is considerably
increased, in excellent agreement with experimental data.

Of course there is room for additional improvements. So far, for
numerical simplicity, exist only applications of this theory with
the parameter set NL3, which has no density dependence in the is
isovector channel. At present there exist more modern parameter
sets with the density dependent meson exchange
\cite{DD-ME1,DD-ME2} which give already on the mean field level
improved results for characteristic properties such as the neutron
skin of neutron-rich nuclei or the density dependence of the
symmetry energy. They should also be implemented in the
theoretical investigations of the type discussed in this article.
A further improvement can be achieved by using a more realistic
pairing force. So far there are only investigations available with
a monopole pairing force. It is therefore highly desirable to
implement in the pairing channel a density dependent zero range
force or the finite range Gogny force. All the calculations
presented here have been done in a discrete basis, i.e. in the
spectral representation of the response equation and therefore the
coupling to the continuum is not taken into account properly so
far. This might have a strong influence on application to light
nuclei and therefore relativistic continuum RPA should be extended
to relativistic continuum QRPA and relativistic continuum QTBA.
Finally, so far ground state correlations have been taken into
account only on the RQRPA level. In the non-relativistic case
there exist investigations going beyond this limitations. They
should be extended also to the relativistic calculations.

\begin{acknowledgments}
Valuable discussions with V. Tselyaev are gratefully acknowledged.
This work has been supported in part by the Bundesministerium
f\"{u}r Bildung und Forschung under project 06 MT 246 and by the
DFG cluster of excellence \textquotedblleft Origin and Structure
of the Universe\textquotedblright(www.universe-cluster.de). E.L.
acknowledge financial support from the Hessian LOEWE initiative 
through the Helmholtz International
Center for FAIR and the Russian Federal Agency of
Education, project No. 2.1.1/4779.
\end{acknowledgments}


\end{document}